\documentclass[pra,showpacs,preprintnumbers,amsmath,amssymb]{revtex4}
\usepackage{amsmath}

\usepackage{graphicx}
\usepackage{dcolumn}
\usepackage{bm}
\usepackage[                   
            pdfstartview=FitH,
            colorlinks, 
            pdfborder=001,   
            linkcolor=blue,
            anchorcolor=blue,
            citecolor=blue,
            urlcolor=blue
            ]{hyperref}
\usepackage{subfigure}
\begin{document}


\title{Remote preparation for single-photon state in two degrees of freedom with hyper-entangled states }

\author {Meiyu Wang$^{1,3}$, Fengli Yan$^{1,3}$,}

\email{flyan@mail.hebtu.edu.cn}
\author {Ting Gao$^{2}$}
\email{gaoting@mail.hebtu.edu.cn}
\affiliation {$^{1}$College of Physics, Hebei Normal University, Shijiazhuang 050024, China\\
$^{2}$ School of Mathematics Science, Hebei Normal University, Shijiazhuang 050024, China\\
$^{3}$ Hebei Key Laboratory of Photophysics Research and Application, Shijiazhuang 050024, China}

\date{\today}
\begin{abstract}
{Remote state preparation (RSP) provides a useful way of transferring quantum information between two distant nodes based on the previously shared entanglement. In this paper, we study RSP of an arbitrary single-photon state in two degrees of freedom (DoFs). Using hyper-entanglement as a shared resource, our first goal is to remotely prepare the single-photon state in polarization and frequency DoFs  and the second one is to reconstruct the single-photon state in polarization and time-bin DoFs. In the RSP process, the sender will rotate the quantum state in each DoF of the photon according to the knowledge of the state to be communicated. By performing a projective measurement on the polarization of the sender's photon, the original single-photon state in two DoFs can be remotely reconstructed at the receiver's quantum systems.  This work demonstrates a novel capability for long-distance quantum communication.}
\end{abstract}
\maketitle

\section {Introduction}

Quantum entanglement is now an indispensable resource for modern information technology, and it plays a vital role in many communication situations and information processing, such as quantum teleportation (QT) \cite{Bennett1}, quantum dense coding \cite{Bennett2}, quantum key distribution \cite{Ekert}, quantum secret sharing \cite{Hillery}, quantum secure direct communication \cite{Long, Deng1}, remote state preparation (RSP) \cite{Lo, Pati, Bennett3}, quantum state transfer \cite{Zhangshou} and so on.  Among these applications, QT and RSP are two typical kinds of quantum procedure that allows one party to transmit a quantum state to another. In QT, the sender Alice can transfer an unknown quantum state to the receiver Bob by relying on the shared entangled state and classical communication.  RSP is usually called ``teleportation of a known quantum state",  which means Alice knows the precise state to be transferred to Bob. On the other hand, QT requires complete Bell-state measurement, which is impossible with linear optics only, while the measurement strategy of RSP is relatively simple and flexible. In recent years, RSP has attracted much attention and various RSP protocols have been proposed \cite{Devetak, Zeng, Berry, Leung, Huang, Kurucz1, Kurucz2, An1, An2, Yang, Chen, ZhangZhou, Jiang, Panigrahi}. Some optical realizations of RSP have also been reported \cite{Laurat, Babichev04, Xiang05, Peters05, Yuan07, Mikami07, Wu10,  Killoran, Neves, Radmark, Ra, Dheur, Jeannic}.

Photons are usually considered as an ideal information carriers in long-distance quantum communication. On the one hand, photons have high-speed transmission and conspicuous
low-noise properties. On the other hand, they have more than one degree of freedom (DoF), such as the polarization, frequency, energy, spatial-mode, time-bin, etc.  These DoFs can be used to construct hyper-entanglement, which refer to the entangled in multiple DoFs of a quantum system simultaneously. Recently, hyper-entanglement becomes a promising quantum resource due to its  high  capacity of channel, and it has extensive applications in quantum information processing, such as Bell-state analysis with hyper-entanglement \cite{Kwiat1,Walborn, Schuck,Barbieri,Sheng,Li,Wang2019}, quantum repeaters \cite{Wang}, entanglement swapping \cite{Sheng, Ren, Li2016}, hype-entanglement concentration \cite{WangHong, LiuJiu}, and so on. So far, experimental progress of hyper-entanglement generation has been reported using different photonic DoFs \cite{Simon,Yabushita,Barreiro2005,Barbieri2005,Rossi,Vallone,Gao,Bhatti,Huber}. In optical systems, the most extensive method to generate a hyper-entangled state is the spontaneous parametric down-conversion (SPDC) process in a nonlinear crystal. When a pump laser beam  shines a $\beta$-barium borate (BBO) crystal, the idler photon and the signal photon are produced probabilistically from the crystal. The SPDC can be divided into type-I and type-II according to the type of crystal phase-matching. In the type-I phase-matching, two SPDC photons have the same polarizations. This type can produce two-photon states that are entangled in time, space, frequency and etc. In the type-II phase-matching, photons are in orthogonal polarization states, and the polarization entangled states are prepared. With the type-I and type-II $\beta$ BBO crystal, hyper-entangled state in two or more DoFs has been demonstrated. For example, in 2005, Barreiro et al. \cite{Barreiro2005} experimentally generated a hyper-entangled state in polarization, spatial mode and time-energy DoFs. In 2009, a six-qubit hyper-entangled state was realized by entangling two photons in polarization and two longitudinal momentum DoFs \cite{Vallone}. Later, Bhatti et al. \cite{Bhatti} presented the experimental setup to produce two-photon polarization-orbital angular momentum entangled states.

With hyper-entanglement, the implementation of QT and RSP for quantum state in multiple DoFs has attached a great deal of interests. By far, there have been several interesting QTs or RSPs. In 2010, Sheng et al. \cite{Sheng} discussed the parallel QT of an arbitrary single-qubit state with polarization-spatial mode hyper-entanglement. In 2015, QT of the composite quantum state of a single photon encoded in both spin and orbital angular momentum has been experimentally realized \cite{Wang15}. In 2018, Li et al. \cite{Li2016} demonstrated the QT using polarization-time-bin hyper-entanglement. For RSP, Barreiro et al. \cite{Barreiro2008} remotely  prepared single-photon states entangled in their spin and orbital angular momentum. Subsequently, they reported the remote preparation of two-qubit hybrid entangled states and vector-polarization beams \cite{Barreiro2010}. In their remote entangled-state preparation protocol, single-photon states are encoded in the photon spin and orbital angular momentum. They reconstructed the states by spin-orbit state tomography and transverse polarization tomography. In 2018, Zhou et al. \cite{Zhou2018} proposed a parallel RSP of a single-qubit state both in the polarization and spatial mode DoFs with linear optical elements. In 2019, Jiao et al. \cite{Jiao} presented a deterministic RSP for an arbitrary single-photon two-qubit hybrid state. They also discussed the RSP for two-qubit hybrid states via linear optical elements with partially hyper-entangled states.

In this paper, we will offer two schemes for parallel preparing a single-photon state via hyper-entangled state. The first parallel RSP protocol is based on the polarization-frequency entanglement and the second one uses the polarization-time-bin entanglement. The polarization is the most popular DoF of the photon for its simple manipulation. The frequency DoF has been used in a series of quantum information schemes because of its stability. And the time-bin DoF is suitable for long-distance quantum communication and fundamental experiments, which can be simply discriminated by the time of arrival. In the two RSP schemes, Alice will rotate the quantum state in each DoF of the photon according to her information of the prepared states. By performing a projective measurement on the polarization of her photon, the original single-photon state in two DoFs can be remotely prepared at the Bob's quantum systems. Our schemes may be of interest not only from a theoretical point of view but also from an experimental one.

\section {Parallel RSP with polarization-frequency hyper-entanglement}

A polarization-frequency hyper-entangled state of a photon pair can be written as
\begin{eqnarray}
|\Psi\rangle_{AB}=\frac{1}{2}(|HH\rangle+|VV\rangle)(|\omega_{1}\omega_{1}\rangle+|\omega_{2}\omega_{2}\rangle)_{AB},
\end{eqnarray}
where the subscripts \emph{A} and \emph{B} denote the two photons sent to the two parties, say Alice and Bob, respectively. $|H\rangle$ and $|V\rangle$ are the horizontal and the vertical polarization states of the photons. $\omega_{1}$ and $\omega_{2}$ are the two frequency modes of the photons. It has been shown that the hyper-entangled state in the polarization and the frequency DoFs can be achieved in experiment \cite{Yabushita, Martin,Chen2017}. Now, Alice's goal is to remotely prepare Bob's qubit in the pure state in both the polarization and the frequency DoFs
\begin{eqnarray}
|\psi\rangle_{B}=(\alpha_{0}|H\rangle+\beta_{0}|V\rangle)(\alpha_{1}|\omega_{1}\rangle+\beta_{1}|\omega_{2}\rangle)_{B}
\end{eqnarray}
with $|\alpha_{0}|^{2}+|\beta_{0}|^{2}=1, ~|\alpha_{1}|^{2}+|\beta_{1}|^{2}=1 $. Alice knows the four parameters completely, but Bob does not know them at all.

\begin{figure}
\begin{center}
\subfigure [] {\includegraphics[width=4in]{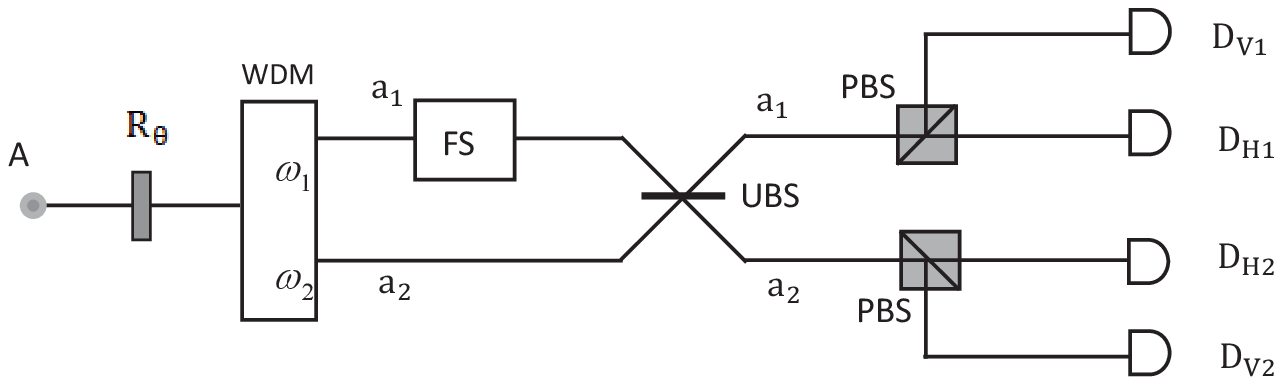}}
\hspace{0.2in}
\subfigure [] {\includegraphics[width=4in]{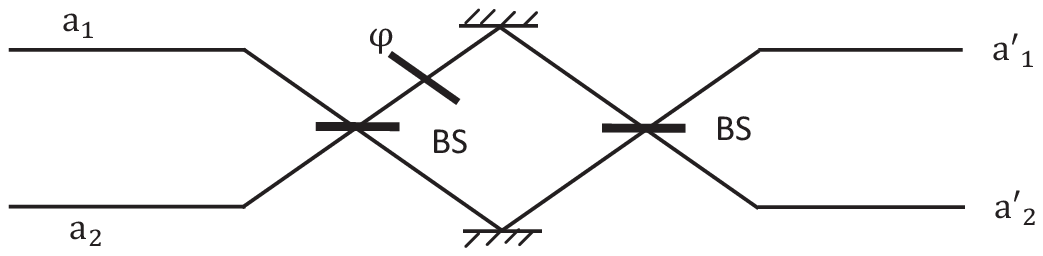}}
\caption{(a) Schematic diagram of the parallel RSP with polarization-frequency hyperentanglement. (b) Schematic diagram of an UBS. BS represents a 50:50 beam splitter. $R_{\theta}$ is a wave plate which can rotate the polarization states of the photon with $\theta$. WDM represents a polarization-independent wavelength division multiplexer which  will lead the photon to different spatial modes according to its frequencies. Frequency shifter (FS) can eliminate the frequency distinguishability. $\varphi$ represents a wave plate which can cause a phase shift between the two spatial modes. PBS represents a polarization beam splitter which transmits the photon in the horizontal polarization state $|H\rangle$ and reflects the photon in the vertical polarization state $|V\rangle$. $D_{i}$ are four single-photon detectors.}
 \end{center}
\end{figure}

The principle of the present parallel RSP using polarization-frequency hyper-entanglement is shown in Fig. 1. First, Alice performs the unitary operation on her hyper-entangled photon on the polarization DoF, which does not perturb the others. She lets the photon \emph{A} pass through a wave plate $R_{\theta}$, which is used to rotate the polarization state
\begin{eqnarray}
|H\rangle &\rightarrow& \mathrm{cos}\theta|H\rangle+ \mathrm{sin}\theta|V\rangle,\nonumber\\
|V\rangle &\rightarrow&  -\mathrm{sin}\theta|H\rangle+\mathrm{cos}\theta|V\rangle
\end{eqnarray}
with $\theta= \mathrm{arccos}\alpha_{0}$. The state $|\Psi\rangle_{AB}$ is transformed to
\begin{eqnarray}
|\Psi\rangle_{1}=\frac{1}{2}[|H\rangle(\alpha_{0}|H\rangle-\beta_{0}|V\rangle)+|V\rangle(\beta_{0}|H\rangle+\alpha_{0}|V\rangle)](|\omega_{1}\omega_{1}\rangle+|\omega_{2}\omega_{2}\rangle)_{AB}.
\end{eqnarray}

Second, Alice performs the unitary operation on her hyper-entangled photon on the frequency DoF, she lets the photon \emph{A} pass through the WDM, FS, UBS sequently. WDM represents a polarization-independent wavelength division multiplexer which is used to divide the photon with frequency $\omega_{1}$ ($\omega_{2}$) into spatial mode $a_{1}$ ($a_{2}$). FS is a frequency shifter with which the frequencies of the photon \emph{A} were converted to the same frequency, e.g., conversion of $\omega_{1}$ to $\omega_{2}$. As a result, we can erase distinguishability for frequency. UBS is an unbalanced beam splitter (See Fig. 1b), which can transform the input state  with spatial modes($a_{1}, a_{2}$) into the output state with spatial modes($a'_{1}, a'_{2}$) \cite{Reck}:
\begin{eqnarray}
\begin{pmatrix}
a'_{1}\\
a'_{2}
\end{pmatrix}
=\begin{pmatrix}
\cos\frac{\varphi}{2} & \sin\frac{\varphi}{2}\\
-\sin\frac{\varphi}{2} & \cos\frac{\varphi}{2}
\end{pmatrix}
\begin{pmatrix}
a_{1}\\
a_{2}
\end{pmatrix}.
\end{eqnarray}
After the WDM, FSs, UBS (with $\frac{\varphi}{2}=\mathrm{arccos}\alpha_{1}$), the evolution of state $|\Psi\rangle_{1}$ can be described as
\begin{eqnarray}
|\Psi\rangle_{1}
&\xrightarrow{\mathrm{WDM}}&\frac{1}{2}[|H\rangle(\alpha_{0}|H\rangle-\beta_{0}|V\rangle)+|V\rangle(\beta_{0}|H\rangle+\alpha_{0}|V\rangle)](|a_{1}\rangle_{\omega_{1}}|\omega_{1}\rangle+|a_{2}\rangle_{\omega_{2}}|\omega_{2}\rangle)_{AB}\nonumber\\
&\xrightarrow{\mathrm{FS}}&\frac{1}{2}[|H\rangle(\alpha_{0}|H\rangle-\beta_{0}|V\rangle)+|V\rangle(\beta_{0}|H\rangle+\alpha_{0}|V\rangle)](|a_{1}\rangle|\omega_{1}\rangle+|a_{2}\rangle|\omega_{2}\rangle)_{AB}\nonumber\\
&\xrightarrow{\mathrm{UBS}}&\frac{1}{2}[|H\rangle(\alpha_{0}|H\rangle-\beta_{0}|V\rangle)(\alpha_{1}|a_{1}\rangle+\beta_{1}|a_{2}\rangle)|\omega_{1}\rangle+|H\rangle(\alpha_{0}|H\rangle-\beta_{0}|V\rangle)(-\beta_{1}|a_{1}\rangle+\alpha_{1}|a_{2}\rangle)|\omega_{2}\rangle\nonumber\\
&&+|V\rangle(\beta_{0}|H\rangle+\alpha_{0}|V\rangle)(\alpha_{1}|a_{1}\rangle+\beta_{1}|a_{2}\rangle)|\omega_{1}\rangle+|V\rangle(\beta_{0}|H\rangle+\alpha_{0}|V\rangle)(-\beta_{1}|a_{1}\rangle+\alpha_{1}|a_{2}\rangle)|\omega_{2}\rangle]\nonumber\\
&=&\frac{1}{2}[|H\rangle|a_{1}\rangle(\alpha_{0}|H\rangle-\beta_{0}|V\rangle)(\alpha_{1}|\omega_{1}\rangle-\beta_{1}|\omega_{2}\rangle)+
|H\rangle|a_{2}\rangle(\alpha_{0}|H\rangle-\beta_{0}|V\rangle)(\beta_{1}|\omega_{1}\rangle+\alpha_{1}|\omega_{2}\rangle)\nonumber\\
&&+|V\rangle|a_{1}\rangle(\beta_{0}|H\rangle+\alpha_{0}|V\rangle)(\alpha_{1}|\omega_{1}\rangle-\beta_{1}|\omega_{2}\rangle)+
|V\rangle|a_{2}\rangle(\beta_{0}|H\rangle+\alpha_{0}|V\rangle)(\beta_{1}|\omega_{1}\rangle+\alpha_{1}|\omega_{2}\rangle).
\end{eqnarray}

The last step is to perform single-photon projective measurement on the photon \emph{A}, and then, she communicates to Bob the outcome of her measurement through a classical channel. With this information, Bob may have to perform a unitary transformation to recover the target state given by Eq. (2). For example, if Alice's measurement result is $|H\rangle_{a_{1}}$, the photon \emph{B} will be in the state $(\alpha_{0}|H\rangle-\beta_{0}|V\rangle)(\alpha_{1}|\omega_{1}\rangle-\beta_{1}|\omega_{2}\rangle)$. Bob can reconstruct the original state  on his entangled photon \emph{B} by performing the corresponding unitary transformation $\sigma_{z}^{p}\otimes\sigma_{z}^{f}$. The relationship between the detection results of the photon \emph{A}, the collapsed states of the photon \emph{B} and the local unitary operations on photon \emph{B} required is shown in table I, according to which one can remotely prepare a single-photon state in polarization and frequency DoFs.
\begin{table}
\centering
\caption{\label{relation} The relation between measurement results of the photon \emph{A}, the final state of \emph{B}, and the local operations.}
\begin{tabular}{ccc}
\hline\hline
Dection (\emph{A}) &state of \emph{B} & local operation \\\hline
$|H\rangle_{a_{1}}$ & $ (\alpha_{0}|H\rangle-\beta_{0}|V\rangle)(\alpha_{1}|\omega_{1}\rangle-\beta_{1}|\omega_{2}\rangle)$ &$\sigma_{z}^{p}\otimes\sigma_{z}^{f}$\\
$|H\rangle_{a_{2}}$ & $ (\alpha_{0}|H\rangle-\beta_{0}|V\rangle)(\beta_{1}|\omega_{1}\rangle+\alpha_{1}|\omega_{2}\rangle)$ &$\sigma_{z}^{p}\otimes\sigma_{x}^{f}$\\
$|V\rangle_{a_{1}}$ & $ (\beta_{0}|H\rangle+\alpha_{0}|V\rangle)(\alpha_{1}|\omega_{1}\rangle-\beta_{1}|\omega_{2}\rangle)$ &$\sigma_{x}^{p}\otimes\sigma_{z}^{f}$\\
$|V\rangle_{a_{2}}$ & $ (\beta_{0}|H\rangle+\alpha_{0}|V\rangle)(\beta_{1}|\omega_{1}\rangle+\alpha_{1}|\omega_{2}\rangle)$ &$\sigma_{x}^{p}\otimes\sigma_{x}^{f}$\\
\hline\hline
\end{tabular}
\end{table}

\section {Parallel RSP with polarization-time-bin hyper-entanglement}

The time-bin DoF is also a simple, conventional DoF. Two different times of arrival can be used to encode the logical 0 and 1. Furthermore,
time-bin entanglement has been proven suitable for long-distance applications, especially in optical fibers \cite{Thew,Marcikic,Inagaki,Valivarthi,Sun}. In
2002, Thew et al. \cite{Thew} reported robust time-bin qubits for distributed quantum communication over 11 km. In 2004, Marcikic et al. \cite{Marcikic} demonstrated the distribution of time-bin entangled qubits over 50 km of optical fiber. Recently, Ref. \cite{Inagaki} reported the distribution of time-bin entangled photon pairs
over 300 km of optical fiber. Also, time-bin encodings have been used in teleportation across fibre networks \cite{Valivarthi,Sun}.

A polarization-time-bin hyper-entangled state of a photon pair can be written as
\begin{eqnarray}
|\Phi\rangle_{AB}=\frac{1}{2}(|HH\rangle+|VV\rangle)(|ee\rangle+|ll\rangle)_{AB},
\end{eqnarray}
where \emph{e} and \emph{l} represent the early and late arrival time of the photons, respectively. The preparation of time-bin entangled states \cite{Schuck, Brendel, Simon1, Barreiro, Zavatta, Barreiro1} have been widely discussed and it can be achieved in experiment.

Suppose that Alice wants to remotely prepare Bob's qubit in the single-photon state in both the polarization and the time-bin DoFs
\begin{eqnarray}
|\phi\rangle_{B}=(\alpha_{0}|H\rangle+\beta_{0}|V\rangle)(\alpha_{2}|e\rangle+\beta_{2}|l\rangle)_{B}
\end{eqnarray}
with $|\alpha_{0}|^{2}+|\beta_{0}|^{2}=1, ~|\alpha_{2}|^{2}+|\beta_{2}|^{2}=1 $.  Alice knows the coefficients $\alpha_{0}, ~\alpha_{2},~\beta_{0},~\beta_{2}$ completely, but Bob does not know them at all.

The device of this RSP of the single-photon two-qubit state is shown in Fig. 2. Similar to the case with the polarization-frequency hyper-entanglement, the first step prepares the polarization state and the second step one deal with the time-bin state. When the photon \emph{A} passes through the wave plate $R_{\theta}$ with $\theta= \mathrm{arccos}\alpha_{0}$. The state $|\Phi\rangle_{AB}$ is transformed to
\begin{eqnarray}
|\Phi\rangle_{1}=\frac{1}{2}(|H\rangle|\eta\rangle+|V\rangle|\eta_{\perp}\rangle)(|ee\rangle+|ll\rangle)_{AB},
\end{eqnarray}
where $|\eta\rangle=\alpha_{0}|H\rangle-\beta_{0}|V\rangle, |\eta_{\perp}\rangle=\beta_{0}|H\rangle+\alpha_{0}|V\rangle$.

Then, Alice performs the unitary operation on her hyper-entangled photon on the temporal DoF, she lets the photon \emph{A} pass through the $\mathrm{PBS_{1}}$, PCs, $\mathrm{PBS_{2}}$, UIs and $R_{\theta_{i}}$ sequently. Here the Pockels cell (PC) \cite{Kalamidas} can flip the polarizations of a photon at a specific time, i.e., the $\mathrm{PC}_{l}$ ($\mathrm{PC}_{e}$) is activated only when the \emph{l} (\emph{e}) component is present.  Each unbalanced interferometer (UI) composed of a pair of PBSs is used to adjust the time-bin state such that the path length difference between the long arm \emph{l} and the short one \emph{e} cancels the time difference between the two time-bins. The wave plates $R_{\theta_{1}}, R_{\theta_{2}}$ can transform the polarization states of the photon \emph{A} to the following state:
\begin{eqnarray}
|H\rangle &\xrightarrow{R_{\theta_{1}}}& \alpha_{2}|H\rangle+\beta_{2}|V\rangle,\nonumber\\
|V\rangle &\xrightarrow{R_{\theta_{2}}}& \alpha_{2}|H\rangle+\beta_{2}|V\rangle
\end{eqnarray}
with $\theta_{1}= \mathrm{arccos}\alpha_{2}, ~\theta_{2}= \mathrm{arccos}\alpha_{2}-\frac{\pi}{2}$. After $\mathrm{PBS_{1}}$, PCs, $\mathrm{PBS_{2}}$, UIs and $R_{\theta_{i}}$, the state evolves as

\begin{eqnarray}
|\Phi_{2}\rangle
&=&\frac{1}{2}(|H^{e}\rangle|\eta^{e}\rangle+|H^{l}\rangle|\eta^{l}\rangle+|V^{e}\rangle|\eta_{\perp}^{e}\rangle+|V^{l}\rangle|\eta_{\perp}^{l}\rangle)_{AB}\nonumber\\
&\xrightarrow{\mathrm{PBS_{1}}}&\frac{1}{2}(|H^{e}\rangle_{a_{2}}|\eta^{e}\rangle_{B}+|H^{l}\rangle_{a_{2}}|\eta^{l}\rangle_{B}+|V^{e}\rangle_{a_{1}}|\eta_{\perp}^{e}\rangle_{B}+|V^{l}\rangle_{a_{1}}|\eta_{\perp}^{l}\rangle_{B})\nonumber\\
&\xrightarrow{\mathrm{PCs}}&\frac{1}{2}(|V^{e}\rangle_{a_{2}}|\eta^{e}\rangle_{B}+|H^{l}\rangle_{a_{2}}|\eta^{l}\rangle_{B}+|V^{e}\rangle_{a_{1}}|\eta_{\perp}^{e}\rangle_{B}+|H^{l}\rangle_{a_{1}}|\eta_{\perp}^{l}\rangle_{B})\nonumber\\
&\xrightarrow{\mathrm{PBS_{2}}}&\frac{1}{2}(|V^{e}\rangle_{a_{2}}|\eta^{e}\rangle_{B}+|H^{l}\rangle_{a_{1}}|\eta^{l}\rangle_{B}+|V^{e}\rangle_{a_{1}}|\eta_{\perp}^{e}\rangle_{B}+|H^{l}\rangle_{a_{2}}|\eta_{\perp}^{l}\rangle_{B})\nonumber\\
&\xrightarrow[R_{\theta_{i}}]{\mathrm{UIs}}&\frac{1}{2}(\alpha_{2}|H^{el}\rangle_{k_{3}}|\eta^{e}\rangle_{B}+\beta_{2}|V^{el}\rangle_{k_{4}}|\eta^{e}\rangle_{B}+\alpha_{2}|H^{le}\rangle_{k_{1}}|\eta^{l}\rangle_{B}+\beta_{2}|V^{le}\rangle_{k_{2}}|\eta^{l}\rangle_{B}\nonumber\\
&&+\alpha_{2}|H^{el}\rangle_{k_{2}}|\eta_{\perp}^{e}\rangle_{B}+\beta_{2}|V^{el}\rangle_{k_{1}}|\eta_{\perp}^{e}\rangle_{B}+\alpha_{2}|H^{le}\rangle_{k_{4}}|\eta_{\perp}^{l}\rangle_{B}+\beta_{2}|V^{le}\rangle_{k_{3}}|\eta_{\perp}^{l}\rangle_{B})\nonumber\\
&=&|\Phi_{3}\rangle.
\end{eqnarray}
We find  that the time difference between the two time-bins of the photon \emph{A} is cancelled, so we omit it in the subsequent expressions.

\begin{figure}
\begin{center}
\scalebox{0.8}{\includegraphics* {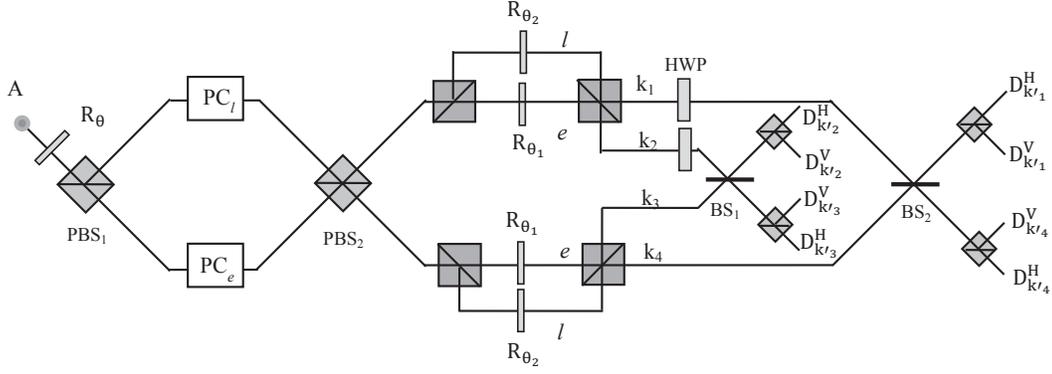}}
\caption{Schematic diagram of the setup for parallel RSP with polarization-time-bin hyperentanglement. PC is a Pockels cell which is used to perform polarization bit-flip operation corresponding to arrival time of the photon. Each pair of PBSs is used to build an unbalanced interferometer (UI). The length difference between the two arms of PBSs in each UI is to cancel the time interval between the two time-bins. $R_{\theta_{1}},R_{\theta_{2}}$ represent the wave plates which rotate the polarization states with angles $\theta_{1}=\mathrm{arccos}\alpha_{2}$, $\theta_{2}=\mathrm{arccos}\alpha_{2}-\frac{\pi}{2}$.  A half-wave plate (HWP) is used to perform the operation $|H\rangle\rightleftharpoons|V\rangle$.}
\end{center}
\end{figure}

Subsequently, two half-wave plates (HWPs) are inserted into  the paths $k_{1}$ and $k_{2}$ respectively, which complete a polarization  bit-flip operation $|H\rangle\rightleftharpoons|V\rangle$. Before the photon \emph{A} is measured, Alice lets it pass through the beam splitters $\mathrm{BS_{1},~BS_{2}}$.  According to the following rules, $a^{\dagger}_{k_{1}}\rightarrow \frac{1}{\sqrt{2}}(a^{\dagger}_{k'_{1}}+a^{\dagger}_{k'_{4}})$, $ a^{\dagger}_{k_{4}}\rightarrow \frac{1}{\sqrt{2}}(a^{\dagger}_{k'_{1}}-a^{\dagger}_{k'_{4}})$, $a^{\dagger}_{k_{2}}\rightarrow \frac{1}{\sqrt{2}}(a^{\dagger}_{k'_{2}}+a^{\dagger}_{k'_{3}})$, $a^{\dagger}_{k_{3}}\rightarrow \frac{1}{\sqrt{2}}(a^{\dagger}_{k'_{2}}-a^{\dagger}_{k'_{3}})$, the state $|\Phi_{3}\rangle$ evolves as follows

\begin{eqnarray}
|\Phi_{3}\rangle
&=&\frac{1}{2}(\alpha_{2}|H\rangle_{k_{3}}|\eta^{e}\rangle_{B}+\beta_{2}|V\rangle_{k_{4}}|\eta^{e}\rangle_{B}+\alpha_{2}|H\rangle_{k_{1}}|\eta^{l}\rangle_{B}+\beta_{2}|V\rangle_{k_{2}}|\eta^{l}\rangle_{B}\nonumber\\
&&+\alpha_{2}|H\rangle_{k_{2}}|\eta_{\perp}^{e}\rangle_{B}+\beta_{2}|V\rangle_{k_{1}}|\eta_{\perp}^{e}\rangle_{B}+\alpha_{2}|H\rangle_{k_{4}}|\eta_{\perp}^{l}\rangle_{B}+\beta_{2}|V\rangle_{k_{3}}|\eta_{\perp}^{l}\rangle_{B})\nonumber\\
&\xrightarrow{\mathrm{HWP}}&\frac{1}{2}(\alpha_{2}|H\rangle_{k_{3}}|\eta^{e}\rangle_{B}+\beta_{2}|V\rangle_{k_{4}}|\eta^{e}\rangle_{B}+\alpha_{2}|V\rangle_{k_{1}}|\eta^{l}\rangle_{B}+\beta_{2}|H\rangle_{k_{2}}|\eta^{l}\rangle_{B}\nonumber\\
&&+\alpha_{2}|V\rangle_{k_{2}}|\eta_{\perp}^{e}\rangle_{B}+\beta_{2}|H\rangle_{k_{1}}|\eta_{\perp}^{e}\rangle_{B}+\alpha_{2}|H\rangle_{k_{4}}|\eta_{\perp}^{l}\rangle_{B}+\beta_{2}|V\rangle_{k_{3}}|\eta_{\perp}^{l}\rangle_{B})\nonumber\\
&\xrightarrow[\mathrm{PBSs}]{\mathrm{BSs}}&\frac{1}{2\sqrt{2}}[|H\rangle_{k'_{1}}(\beta_{0}|H\rangle+\alpha_{0}|V\rangle)(\beta_{2}|e\rangle+\alpha_{2}|l\rangle)_{B}+|H\rangle_{k'_{2}}(\alpha_{0}|H\rangle-\beta_{0}|V\rangle)(\alpha_{2}|e\rangle+\beta_{2}|l\rangle)_{B}\nonumber\\
&&+|H\rangle_{k'_{3}}(\alpha_{0}|H\rangle-\beta_{0}|V\rangle)(-\alpha_{2}|e\rangle+\beta_{2}|l\rangle)_{B}+|H\rangle_{k'_{4}}(\beta_{0}|H\rangle+\alpha_{0}|V\rangle)(\beta_{2}|e\rangle-\alpha_{2}|l\rangle)_{B}\nonumber\\
&&+|V\rangle_{k'_{1}}(\alpha_{0}|H\rangle-\beta_{0}|V\rangle)(\beta_{2}|e\rangle+\alpha_{2}|l\rangle)_{B}+|V\rangle_{k'_{2}}(\beta_{0}|H\rangle+\alpha_{0}|V\rangle)(\alpha_{2}|e\rangle+\beta_{2}|l\rangle)_{B}\nonumber\\
&&+|V\rangle_{k'_{3}}(\beta_{0}|H\rangle+\alpha_{0}|V\rangle)(\alpha_{2}|e\rangle-\beta_{2}|l\rangle)_{B}+|V\rangle_{k'_{4}}(\alpha_{0}|H\rangle-\beta_{0}|V\rangle)(-\beta_{2}|e\rangle+\alpha_{2}|l\rangle)_{B}].
\end{eqnarray}

Obviously, the photon \emph{A} will emit from different exits of the setups and trigger different single-photon detectors. The relationship between the detection results of the photon \emph{A}, the collapsed states of the photon \emph{B} and the local unitary operations on photon \emph{B} required is shown in table II, according to which one can remotely prepare a single-photon state in polarization and time-bin DoFs.

\begin{table}
\centering
\caption{\label{relation} The relation between measurement results of the photon \emph{A}, the final state of \emph{B}, and the local operations.}
\begin{tabular}{ccc}
\hline\hline
Dection(\emph{A}) &state of \emph{B} & local operation \\\hline
$|H\rangle_{k'_{1}}$ & $ (\beta_{0}|H\rangle+\alpha_{0}|V\rangle)(\beta_{2}|e\rangle+\alpha_{2}|l\rangle)$ &$\sigma_{x}^{p}\otimes\sigma_{x}^{t}$\\
$|H\rangle_{k'_{2}}$ & $ (\alpha_{0}|H\rangle-\beta_{0}|V\rangle)(\alpha_{2}|e\rangle+\beta_{2}|l\rangle)$ &$\sigma_{z}^{p}\otimes I $\\
$|H\rangle_{k'_{3}}$ & $ (\alpha_{0}|H\rangle-\beta_{0}|V\rangle)(-\alpha_{2}|e\rangle+\beta_{2}|l\rangle)$ &$\sigma_{z}^{p}\otimes\sigma_{z}^{t}$\\
$|H\rangle_{k'_{4}}$ & $ (\beta_{0}|H\rangle+\alpha_{0}|V\rangle)(\beta_{2}|e\rangle-\alpha_{2}|l\rangle)$ &$\sigma_{x}^{p}\otimes i\sigma_{y}^{t}$\\
$|V\rangle_{k'_{1}}$ & $ (\alpha_{0}|H\rangle-\beta_{0}|V\rangle)(\beta_{2}|e\rangle+\alpha_{2}|l\rangle)$ &$\sigma_{z}^{p}\otimes\sigma_{x}^{t}$\\
$|V\rangle_{k'_{2}}$ & $ (\beta_{0}|H\rangle+\alpha_{0}|V\rangle)(\alpha_{2}|e\rangle+\beta_{2}|l\rangle)$ & $\sigma_{x}^{p}\otimes I$\\
$|V\rangle_{k'_{3}}$ & $ (\beta_{0}|H\rangle+\alpha_{0}|V\rangle)(\alpha_{2}|e\rangle-\beta_{2}|l\rangle)$ & $\sigma_{x}^{p}\otimes\sigma_{z}^{t}$\\
$|V\rangle_{k'_{4}}$ & $ (\alpha_{0}|H\rangle-\beta_{0}|V\rangle)(-\beta_{2}|e\rangle+\alpha_{2}|l\rangle)$ & $\sigma_{z}^{p}\otimes i\sigma_{y}^{t}$\\
\hline\hline
\end{tabular}
\end{table}

\section{Discussion and conclusion }

It is interesting to compare our parallel RSP with the previous parallel RSP in Ref. \cite{Zhou2018}. Similar to Refs. \cite{Jiang2012,Deng2017, Zhou2018}, the efficiency of RSP can be defined by the formula $\eta=\frac{q_{s}}{q_{u}+b_{t}}$ with $q_{s}$ denoting the number of qubits transmitted, $q_{u}$ denoting the number of the qubits that is used as the quantum channel, and  $b_{t}$ representing the number of classical bit needed. On the one hand, these RSP protocols' goals are all to remotely prepare a single-photon two-qubit state via the hyper-entanglement Bell state in two DoFs, so $q_{s}=2$  and $q_{u}=4$ for each protocol. On the other hand, in Ref. \cite{Zhou2018}, the single-photon state in polarization and spatial-mode DoFs can be remotely prepared  with cost of 3 cbits. Our first RSP scheme for the single-photon state in polarization and frequency DoFs needs only 2 cbits.  So the efficiencies of the Ref. \cite{Zhou2018} and our first protocol are $\frac{2}{7}$ and $\frac{1}{3}$ respectively. Obviously, the latter's efficiency is higher than that of the former.  However, our first protocol exploits WDM to transform the frequency DoF of the photon into its spatial DoF. With FS, the distinguishability for the frequency is erased. So this is a trade-off between our first RSP and  Ref. \cite{Zhou2018}.  For our second RSP of the single-photon state both in polarization DoF and time-bin DoFs, it can be completed with the same classical information as in Ref. \cite{Zhou2018}, that is, the efficiency of our second protocol is the same as that of \cite{Zhou2018}.  But in our second RSP, it only exploits a fiber channel for each party. Using the time-bin DoF rather than the spatial-mode DoF requires less resources in long-distance quantum communication.

Now let us discuss the possibility in experiment. In our protocols for RSP of arbitrary single-photon state in two DoFs, linear optical elements such as PBSs, BSs, wave plates, HWPs, and PCs are used to perform operations on the photon. In addition, the optical devices WDM and FS are necessary for performing operations on the frequencies of the photon.  Current available PC can be meet the requirement for the order of a few nanoseconds. In particular,  Zhang et al. \cite{Zhang4} showed that the switch efficiency of plasmsa PC in whole aperture is better than $99\%$. The WDM, which is used to lead photons into different spatial modes according to their different frequencies, can be realized with optical cavity \cite{Huntington1,Zhang1}, asymmetric Mach-Zehnder on the frequency encoding \cite{Huntington2,Huntington3}, and Fiber Bragg Grating \cite{Bloch,Zhang2}. The FS, which is used to eliminate the frequency distinguishability, can be implemented by means of frequency up-conversion process or down-conversion process \cite{Takesue05, Takesue08, Ikuta11, Zhou16}. All of these show the operations required in our protocols can be implemented with current technology. However, we assume the efficiency of the optical elements to be perfect, i.e., there is no photon loss in these optical elements. Also, the efficiency of the single-photon detectors is assumed to be $100\%$. In practice, they do not work ideally. For instance, the low  detection efficiency caused by photon loss has an effect on the efficiency of our RSP protocols. When single-photon detectors of a finite efficiency $\eta_{d}$ are used, the efficiency of our RSP protocols would be decreased by a scale of $\eta^{2}_{d}$. In this case, the post-selection is necessary.

In summary, we have proposed two parallel RSP protocols for  arbitrary single-photon two-qubit state based on hyper-entanglement. The first protocol is to remotely prepare the single-photon state in polarization and frequency DoFs with the polarization-frequency entanglement. The second one is to prepare the single-photon state in polarization and time-bin DoFs with the polarization-time-bin entanglement.  In the two RSP schemes, resorting to optical elements, the sender will rotate the quantum state in each DoF of the photon according to the information of the prepared states. By performing a projective measurement on the polarization of the sender's photon, the original single-photon state in two DoFs can be remotely reconstructed at the receiver's quantum systems. Our protocols are useful for long-distance quantum communication.

\vspace{0.5cm}

{\noindent\bf Acknowledgments}\\[0.2cm]

This work was supported by the National Natural Science Foundation of China under Grant Nos. 11475054 and 12071110, Hebei Natural Science Foundation of China under Grant Nos. A2019205190, A2020205014 and A2018205125, Graduate Scientific Innovative Foundation of the Education Department of Hebei Province under Grant No.CXZZBS2019079, the Education Department of Hebei Province Natural Science Foundation under Grant No. ZD2020167, and Science Foundation of Hebei Normal University under Grant No. L2021B13.


\vspace{0.5cm}

\end{document}